\documentclass [12pt]{article}
\usepackage{amsmath}
\usepackage{amsfonts}
\usepackage{epsfig}
\begin{document}
\def\b{\bar}
\def\d{\partial}
\def\D{\Delta}
\def\cD{{\cal D}}
\def\cK{{\cal K}}
\def\f{\varphi}
\def\g{\gamma}
\def\G{\Gamma}
\def\l{\lambda}
\def\L{\Lambda}
\def\M{{\Cal M}}
\def\m{\mu}
\def\n{\nu}
\def\p{\psi}
\def\q{\b q}
\def\r{\rho}
\def\t{\tau}
\def\x{\phi}
\def\X{\~\xi}
\def\~{\widetilde}
\def\h{\eta}
\def\bZ{\bar Z}
\def\cY{\bar Y}
\def\bY3{\bar Y_{,3}}
\def\Y3{Y_{,3}}
\def\z{\zeta}
\def\Z{{\b\zeta}}
\def\Y{{\bar Y}}
\def\cZ{{\bar Z}}
\def\`{\dot}
\def\be{\begin{equation}}
\def\ee{\end{equation}}
\def\bea{\begin{eqnarray}}
\def\eea{\end{eqnarray}}
\def\half{\frac{1}{2}}
\def\fn{\footnote}
\def\bh{black hole \ }
\def\cL{{\cal L}}
\def\cH{{\cal H}}
\def\cF{{\cal F}}
\def\cP{{\cal P}}
\def\cM{{\cal M}}
\def\ik{ik}
\def\mn{{\mu\nu}}
\def\a{\alpha}

\title{Gravity versus Quantum theory:\\ Is electron really  pointlike?}
\author{Alexander Burinskii \\
Theor.Physics Laboratory, NSI, Russian Academy of Sciences,\\ B.
Tulskaya 52 Moscow 115191 Russia, email: bur@ibrae.ac.ru}

\date{Essay written for the Gravity Research Foundation 2011
Awards for Essays on Gravitation. (March 31, 2011) } \maketitle

\begin{abstract}
Quantum theory claims that electron is pointlike and
structureless. Contrary,  the consistent with Gravity Kerr-Newman
(KN) electron model displays an extended structure of the Compton
size $r_c=\hbar /m .$ We obtain that there is no real conflict
between the extended Gravitating electron and a Quantum electron
"dressed" by virtual particles. In the same time the KN model
indicates new important details of the electron structure and
sheds new light on some old puzzles of quantum theory. In
particular, the KN Gravity predicts that electron forms a disklike
vacuum bubble bounded by a closed string, which could probably be
detected by the novel experiments. If it will be confirmed, it
would be of primary importance for foundations of Quantum theory
and unification of Quantum theory with Gravity.
\end{abstract}

\newpage

\hfill {\it "Nobody understands quantum mechanics." }

\hfill Richard Feynman (1965), \cite{RFeyn}

\vspace{0.4cm}

Modern physics is based on Quantum theory and Gravity. The both
theories are confirmed experimentally with great precision.
Nevertheless, they are conflicting and cannot be unified in a
whole theory. In this essay we discuss one of the principal
contradictions, the question on the shape and size of
 electron.

Quantum theory states that electron is pointlike and
structureless. In particular,
 Frank Wilczek  writes in \cite{FWil}: "...There's no evidence that electrons have
internal structure (and a lot of evidence against it)", while the
superstring theorist Leonard Susskind notes
 that  electron radius is "...most probably not much bigger and
not much smaller than the Planck length..",   \cite{LSuss}.
\fn{Author thanks Don Stevens for these references and
conversation.} This point of view is supported by experimental
evidences, which have not found the electron structure down to
$10^{-16} cm .$

The widespread opinion that the range of interaction for
gravitational field is "tremendously weak" and becomes compatible
to other forces only at Planck scale, \cite{tHoof}, is inspired by
the Schwarzschild relation $ r_g=2m .$
 The Kerr geometry turns this relation into inverse one, $r_g \sim J/m,$
 which points out that the range of interaction may be extended to
 radius of the Kerr singular ring, $ a = J/m .$
 Gravitational field of the Kerr solution concentrates in a
 thin vicinity of the Kerr ring, forming a type of ``gravitational
 waveguide'', or string. For electron,  the Kerr field may be extended to
  the Compton radius $r_c=\hbar/(2m),$ which corresponds to the size of
  a "dressed" electron. We argue here that the Kerr string is an element
  of the extended electron structure.

 In 1968 Carter obtained that the KN solution for the charged and
 rotating black holes has $g=2$ as that of the Dirac electron, \cite{Car,DKS},
which initiated development of the electron models  based on the
KN solution
\cite{Car,DKS,Isr,BurGeonIII,Bur0,Bur01,IvBur,Lop,BurSen,BurStr,BurTwi,
BurBag,Arc,Dym,TN,BurSol}.

In the units $c=\hbar =G=1 , $ mass of electron is $m\approx
10^{-22},$ while $ \ a=J/m \approx 10^{22} .$ Therefore, $a>>m ,$
and the black hole horizons disappear, opening the Kerr singular
ring which is a branch line of the twovalued Kerr spacetime.
Development of the KN electron models for four decades formed
severe lines of investigation:

\begin{figure}[ht]
\centerline{\epsfig{figure=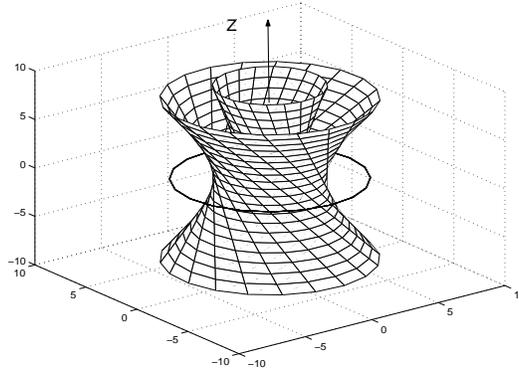,height=5cm,width=7cm}}
\caption{Vortex of the Kerr congruence. Twistor null lines are
focused on the Kerr singular ring, forming a circular
gravitational waveguide, or string with lightlike excitations.}
\end{figure}

\begin{itemize}

\item[$(a)$] First ("thin shell") model was suggested by Israel,
\cite{Isr}, who truncated the "negative" fold of metric, forming a
rotating disk spanned by the Kerr singular ring. Hamity \cite{Ham}
showed that the disk is rigidly rotating.

\item[$(b)$]  L\'opez \cite{Lop} removed the Kerr singular ring
together with negative fold, forming a rotating disklike bubble
with a flat interior.

\item[$(c)$] "Microgeon" models
\cite{BurGeonIII,Bur0,Bur01,BurGeon0} evolved into $4D$ string
models
\cite{IvBur,BurSen,BurStr,BurTwi,TN,BurOri,Nish,BurAli,BurKN,BurAxi}.

\item[$(d)$] Superconducting bag models \cite{BurBag,BEHM} based
either on  nonlinear electrodynamics \cite{Dym,BurHild}, or on the
Higgs field model \cite{BurBag,RenGra}

\item[$(e)$] Gravitating soliton model \cite{BurSol} is
development of the type $(c)$ and $(d)$ models.

\end{itemize}

All these models unambiguously indicated  Compton radius of the
electron. Note, that the Compton radius plays also peculiar role
in the Dirac theory, as a limit of localization of the wave
packet. Localization beyond the Compton zone creates a
"zitterbewegung" affecting "...such paradoxes and dilemmas, which
cannot be resolved in frame of the Dirac electron theory..."
(Bjorken and Drell, \cite{BjoDr}). Dirac wrote in his Nobel Prize
Lecture : "The variables $\alpha$ (velocity operators, AB) also
give rise to some rather unexpected phenomena concerning the
motion of the electron. .. It is found that an electron which
seems to us to be moving slowly, must actually have a very high
frequency oscillatory motion of small amplitude superposed on the
regular motion which appears to us. As a result of this
oscillatory motion, the velocity of the electron at any time
equals the velocity of light."

 {\bf Mass without mass.}  The puzzle of "zitterbewegung"  and the
 known processes of
 annihilation of the electron-positron pairs brought us in 1971 to
the Wheeler "geon" model of "mass without mass" \cite{Wheel}.
 In \cite{BurGeon0} we considered a massless particle circulating
around z-axis. Its local 4-momentum is lightlike,  \be p_x^2 +
p_y^2 + p_z^2 = E^2 \label{Ephot} ,\ee while the effective
mass-energy was created by an averaged orbital motion,
 \be <p_x^2> +<p_y^2> = \tilde m^2
\label{mPxy} .\ee Averaging (\ref{Ephot}) under the condition
(\ref{mPxy}) yields  \be <p_x^2 + p_y^2 + p_z^2> = \tilde m^2
+p_z^2 = E^2 \label{mPzE} .\ee Quantum analog of this model
corresponds to a wave function $\psi(\vec x,t) $ and operators, $
\vec p \to \hat {\vec p} = -i\hbar \nabla , \quad \hat E= i \hbar
\d_t .$  From (\ref{Ephot}) and (\ref{mPxy}) one obtains two wave
equations: \be (\d_x^2 + \d_y^2)\psi = \tilde m^2  \psi =  (\d_t
^2
 - \d_z^2)\psi \label{msep} ,\ee which may be separated by the ansatz
 \be\psi ={\cal{M}}(x,y)\Psi_0 (z,t)
\label{ans}. \ee  The RHS of (\ref{msep}) yields the usual
equation for a massive particle, $ (\d_t ^2 - \d_z^2)\Psi_0
=\tilde m^2 \Psi_0 ,$ and the corresponding (de Broulie) plane
wave solution \be \Psi_0 (z,t) =\exp{\frac i \hbar (z p_z -Et)} ,
\label{deBr} \ee while the LHS determines the ``internal''
structure factor \be {\cal{M}}_\n={\cal{H}}_\n (\frac {\tilde m}
\hbar \rho)
 \exp \{i\n \phi \} \label{MHan}, \ee
in polar coordinates $\rho, \phi ,$ where
  ${\cal{H}}_\n ( \frac {\tilde m } \hbar \rho) $ are the Hankel
functions of index $\n.$   ${\cal{M}}_\n$ are eigenfunctions of
operator
 $\hat J_z = \frac \hbar i \d_\phi $ with eigenvalues $J_z=
\n\hbar .$ For electron we have $J_z= \pm \hbar/2, \quad \n=\pm
1/2 ,$ and the factor \be {\cal{M}}_{\pm 1/2}= \rho^{-1/2}\exp \{
i (\frac {\tilde m} \hbar \rho \pm \frac 12 \phi )\} \label{M12}
\ee creates a singular ray along $z$-axis, which forms a branch
line, and the wave function is twovalued.

 There exits also the corresponding spinor model \cite{BurGeon0}
 generating Dirac equation from the initially massless one.

{\bf The Kerr string.}  Principal problem of this model was the
weakness of the Schwarzschild gravitational field, strength of
which fails about $22$ orders. The works \cite{Car,DKS,Isr}
appeared as a stunning surprise, which determined all subsequent
development of the type (c) models. The Kerr gravitational field
is concentrated near the Kerr singular ring and forms a {\it
gravitational waveguide} for traveling waves.
 Indeed, it was recognized soon that the Kerr singular ring is a type of
 gravitational string \cite{IvBur,BurSen,BurOri,BurAli}, while the
 traveling waves are stringy excitations.\fn{It
 was shown in \cite{BurSen} that structure of the field
around the Kerr ring is similar to the field around a heterotic
string.} It has been shown that the Kerr metric provides self
-consistency of the spinning geon model \cite{BurGeonIII}. First
approximate solutions were considered in \cite{Bur0}, while the
exact solutions for electromagnetic excitations on the Kerr-Schild
background represented a very hard problem \cite{Kerr} and were
obtained much later \cite{BurAxi,BurA,BurExcit,BurExa}. It has
been shown that any wave excitation  creates some `axial' singular
ray (see Fig.2) similar to the `axial' singular ray of the geon
model.

\begin{figure}[ht]
\centerline{\epsfig{figure=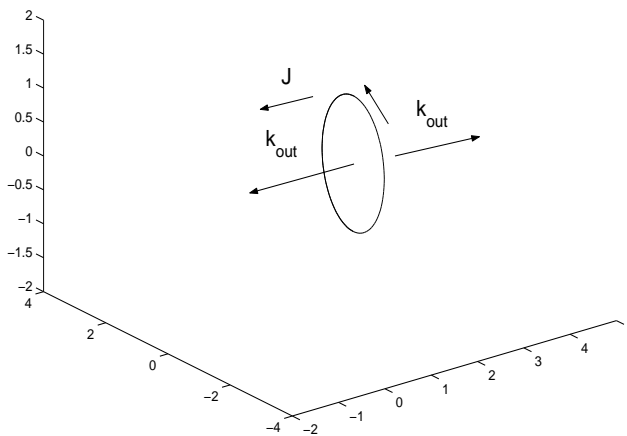,height=7cm,width=8.5cm}}
\caption{Skeleton of the Kerr geometry \cite{BurAxi} formed by the
topologically coupled  "circular" and "axial" strings.}
\end{figure}

{\bf Gravitating KN soliton.} The KN soliton model \cite{BurSol}
represents a field version of the bubble model $(b)$. Surface of
the bubble  is fixed by the Kerr radial coordinate $r=r_e =
e^2/(2m) ,$ and forms an oblate  disk of the Compton radius $r_c
\approx a =\hbar/(2m) .$

Gravitational field is regularized by a chiral field model,
$U(1)\times \tilde U(1),$ which provides a phase transition from
the external KN `vacuum state', $ V_{ext}=0 ,$  to a flat internal
`pseudovacuum' state, $ V_{int}=0 .$ Electromagnetic field is
regularized by the Higgs mechanism of broken symmetry, similarly
to other models of electroweak theory \cite{tHoof,Kus,Mant,Dash}.
 The model exhibits  two essential peculiarities:
\begin{itemize}
\item  the Kerr ring is regularized, forming on the border of
bubble a closed relativistic string of the Compton radius $r_c$
and a quantized loop of electromagnetic potential $ \oint
eA^{(str)}_\phi d\phi=-4\pi ma \label{WL} ,$ which determines
total spin, $J=ma=n/2, \ n=1,2,3,...$,

\item the Higgs field inside the bubble forms a coherent vacuum
state oscillating with frequency $\omega=2m .$

\end{itemize}

The KN soliton forms a regular background  for stringy excitations
described by the type $(c)$ models, while the wave excitations of
the Kerr string are determined by the exact time-dependent
Kerr-Schild solutions, \cite{Bur0,BurAxi,BurA,BurExa}.

{\bf Does the KN model of electron contradict to Quantum Theory?}
It seems ``yes'', if one speaks on the "bare" electron. However,
in accordance with QED, vacuum polarization creates in the Compton
region a cloud of virtual particles forming a "dressed" electron.
This region gives contribution to electron spin, and performs a
procedure of renormalization, which determines physical values of
the electron charge and mass, \cite{BjoDr,Thirr,LifPit2}.
Therefore, speaking on the ``dressed''   electron, one can say
that the real contradiction between the KN model and the Quantum
electron is absent.

 Dynamics of the virtual particles in
QED is chaotic, which allows one to separate conventionally it
from the ``bare''electron. On the other hand,
 the vacuum state inside the KN soliton model forms a {\it coherent
state,} joined with the closed Kerr string. It represents an
`internal' structure which cannot be separated from a ``bare''
particle, but  should be considered as {\it integral whole of the
extended electron}.

We should still comment the absence of experimental exhibitions of
the electron structure. First, it may be caused by a specific
complex structure of the Kerr geometry
\cite{Bur0,BurStr,New,BurCompl,BurNst,BurMag,BurSup}: the KN
solution appears as a real slice of a pointlike source positioned
in complex region.\fn{This representation was obtained by Appel in
1887 \cite{App}. In fact, the KN solution was obtained first in
\cite{New0} by a complex transformation from the
Reissner-Nordstr\"om solution.} Fourier transform of the complex
source is very similar to Fourier image of the real pointlike
source, which may result in its pointlike exhibition in the
momentum space. Alternative explanation (discussed in
\cite{BurTwi}) is related with the lightlike singular beams (see
Fig.2.), accompanying any wave excitation of the Kerr
geometry,\cite{BurKN,BurAxi,BurA,BurExcit}. Finally, the pointlike
interaction may simply be related with the contact character of
the string-string interactions.

{\bf Conclusion:} The KN gravity sheds a new light on the possible
role of Gravity in the structure of Quantum theory. If the
electron has really  the predicted closed string on the boundary
of a disklike bubble, it should apparently be detected
experimentally by a novel effective tool -- the ``nonforward
Compton scattering'' \cite{Rad,Hoyer,Burk,Ji}.

\end{document}